\documentclass[aps,prl,reprint,superscriptaddress,nofootinbib]{revtex4-2}

\usepackage{amsmath,amssymb,bm}
\usepackage{graphicx}
\usepackage{braket}
\usepackage{hyperref}
\usepackage{xcolor}

\begin{document}

\title{Latent Symmetry in Quantum Zeno Dynamics}

\author{Stefano Longhi}
\email{stefano.longhi@polimi.it}
\affiliation{Dipartimento di Fisica, Politecnico di Milano, Piazza Leonardo da Vinci 32, I-20133 Milano, Italy}
\affiliation{IFISC (UIB-CSIC), Instituto de Fisica Interdisciplinar y Sistemas Complejos, E-07122 Palma de Mallorca, Spain}


\begin{abstract}

Quantum Zeno dynamics constrains a continuously monitored system to
measurement-selected subspaces. Here we show that the structure of
such Zeno subspaces can be governed by a \emph{latent symmetry},
namely a symmetry that is absent from the geometry of the underlying
physical network. We consider a minimal circuit-QED network of coupled
microwave resonators subject to quantum-nondemolition (QND) monitoring
of local photon occupations. Two distinguished resonators are
geometrically inequivalent, yet are exactly exchanged by a latent
symmetry of the network. Strong monitoring partitions the
single-excitation Hilbert space into three quantum-Zeno subspaces, two
of which correspond to localization on the distinguished resonators.
Latent symmetry exactly exchanges these two Zeno subspaces and enforces
identical escape and residence statistics despite their different
microscopic environments. In the strong-measurement regime their
residence times grow linearly with the measurement strength and
diverge in the ideal Zeno limit, where the measurement subspaces
become dynamically disconnected. Our results show that latent symmetry
can act as a hidden organizing principle of quantum Zeno dynamics,
imposing exact relations between measurement-selected subspaces that
are geometrically inequivalent in the physical system.

\end{abstract}

 
\maketitle

\section{Introduction}
Symmetry is one of the central organizing principles of physics. It
classifies eigenstates, produces selection rules, and underlies
spontaneous symmetry breaking and phase transitions. In a network
described by a single-excitation Hamiltonian $H$, an ordinary
\emph{geometrical symmetry} is represented by a permutation matrix $P$
satisfying
\begin{equation}
[H,P]=0.
\label{eq:ordinary_symmetry}
\end{equation}
The concepts of \emph{isospectral reduction} and \emph{latent
symmetry}
\cite{LS1,LS2,LS3,LS4,LS4b,LS4c,LS5,LS6,LS7,LS8,LS9,LS10,LS11,
LS12,LS13,LS14,LS15,LS16,LS18}
have considerably broadened this paradigm by showing that exact
symmetry constraints can persist even when no corresponding
geometrical symmetry is visible in the physical network. Revealed
through an appropriate isospectral reduction of the Hamiltonian,
latent symmetries have found applications in a variety of settings,
including flat-band engineering \cite{LS6}, wave-scattering control
\cite{LS11}, robust topological states \cite{LS9,LS10,LS13},
topological Anderson insulators \cite{LS15}, and non-Hermitian exceptional structures \cite{LS7}.
While such recent applications of latent symmetry have largely focused
on spectral and topological phenomena, its role in continuously
monitored quantum dynamics has received comparatively little
attention. A particularly natural setting in which to explore this
question is provided by the quantum Zeno effect and quantum Zeno
dynamics 
\cite{QZ1,QZ1b,QZ2,KofmanKurizki1996,QZ1bis,QZ3,QZ4,QZ4b,QZ5,QZ6,QZ7,QZ1d,QZ8,QZ1e,QZ1f,QZ8b,QZ8bis,QZ9,QZ10,
KofmanKurizki2000,KofmanKurizki2001,QZ11,QZ1g,QZ1h,QZ12,QZ13,QZ14,QZ1i,QZ15}.
Under sufficiently strong or frequent measurement, transitions between
eigenspaces of the measured observable are suppressed. More generally,
the Hilbert space is partitioned into \emph{quantum-Zeno subspaces}
\cite{QZ5,QZ8}, within which the system can retain nontrivial dynamics
while transitions between different subspaces become progressively
inhibited. Quantum Zeno dynamics thus turns the structure of the
measurement eigenspaces into a central element of the system dynamics. 

This raises a natural question: can a symmetry hidden from the
physical architecture nevertheless organize the structure and dynamics
of quantum-Zeno subspaces? In particular, can latent symmetry impose
exact relations between Zeno sectors associated with geometrically
inequivalent physical configurations?

In this work we show that latent symmetry can indeed act as a hidden
organizing principle of quantum Zeno dynamics. Continuous monitoring
creates measurement-selected Zeno subspaces, while latent symmetry can
impose exact relations among them even when the corresponding physical
configurations are not geometrically equivalent. In particular, we
show that two geometrically inequivalent Zeno sectors exchanged by a
latent symmetry possess identical escape and residence statistics.
This provides a dynamical manifestation of latent symmetry that is
specific to the measurement-conditioned evolution rather than to the
geometry of the underlying network.
We illustrate this mechanism using a minimal circuit quantum electrodynamics (QED) network of
four coupled microwave resonators.  Circuit  QED provides a natural platform for observing quantum Zeno dynamics, combining controllable networks of coupled microwave modes
with dispersive readout \cite{QED1,QED2,QED3,QED4,QED5,QED6,QED7}
and access to individual quantum trajectories \cite{QED4,QED8}.
Continuous-measurement and quantum-trajectory descriptions are well
established in this setting \cite{QND1,QND2,QND3}, making the predicted
latent-symmetry constraints directly accessible from measurement
records. In our minimal network design, two distinguished resonators $u$
and $v$ are geometrically inequivalent but become exactly exchange
symmetric under an isospectral reduction of the network. Their photon
occupations are continuously monitored by quantum-nondemolition (QND)
readout. Strong monitoring partitions the single-excitation Hilbert
space into three Zeno subspaces: two one-dimensional sectors associated
with localization on $u$ and $v$, and a two-dimensional auxiliary
sector. The latent symmetry exactly exchanges the first two while
leaving the auxiliary sector invariant as a subspace. Consequently,
despite their different microscopic coupling environments, the two
polarized Zeno sectors exhibit identical escape and residence
statistics. In the strong-monitoring regime, their escape rates display
the characteristic quantum-Zeno suppression, while breaking the latent
symmetry lifts their exact pairing.
Thus the central result is not the occurrence of Zeno localization
itself, but the hidden symmetry constraint relating Zeno subspaces
that are inequivalent in the physical geometry. More broadly, the
interplay between latent symmetry and continuous measurement provides
a route to organizing quantum Zeno dynamics beyond geometrically
symmetric systems and suggests extensions to more complex,
multiphoton, interacting, and many-body monitored networks.

\section{Latent-symmetric circuit-QED network}

\subsection{Model and latent symmetry}

A minimal model displaying latent symmetry is the four-resonator circuit-QED network schematically shown
in Fig.~1(a). Each resonator supports a single microwave mode, and the
resonators are linearly coupled, for example capacitively or through
tunable coupling elements, allowing microwave excitations to hop
between connected resonators \cite{QED5,QED6,QED7}. Two distinguished
resonators, denoted by $u$ and $v$, are indirectly coupled through two
auxiliary resonators $a$ and $b$. The resonance frequencies of $u$ and
$v$ are taken to be equal to $\epsilon_0$, whereas those of the
auxiliary resonators $a$ and $b$ are equal to $\Delta$. We set
$\hbar=1$ throughout.
The hopping amplitudes connecting resonator $u$ to $a$ and $b$ are
denoted by $g_1$ and $g_2$, respectively, whereas $h_1$ and $h_2$
denote the corresponding couplings from resonator $v$. Introducing
bosonic creation and annihilation operators $\hat a_j^\dagger$ and
$\hat a_j$, with $j=u,v,a,b$, the resonator Hamiltonian reads
\begin{align}
\hat H_0={}&
\epsilon_0(\hat n_u+\hat n_v)
+\Delta(\hat n_a+\hat n_b)
\nonumber\\
&+g_1(\hat a_u^\dagger\hat a_a+\mathrm{H.c.})
+g_2(\hat a_u^\dagger\hat a_b+\mathrm{H.c.})
\nonumber\\
&+h_1(\hat a_v^\dagger\hat a_a+\mathrm{H.c.})
+h_2(\hat a_v^\dagger\hat a_b+\mathrm{H.c.}),
\label{eq:Hsecond}
\end{align}
where
\begin{equation}
\hat n_j=\hat a_j^\dagger\hat a_j
\end{equation}
is the microwave-photon number operator of resonator $j$. The
Hamiltonian conserves the total number of microwave excitations,
\begin{equation}
\hat N=\sum_{j=u,v,a,b}\hat n_j,
\qquad
[\hat H_0,\hat N]=0.
\label{eq:number_conservation}
\end{equation}
The QND monitoring introduced below also conserves $\hat N$, so that
each fixed-excitation sector evolves independently.

\begin{figure*}[t]
\centering
\includegraphics[width=0.95\textwidth]{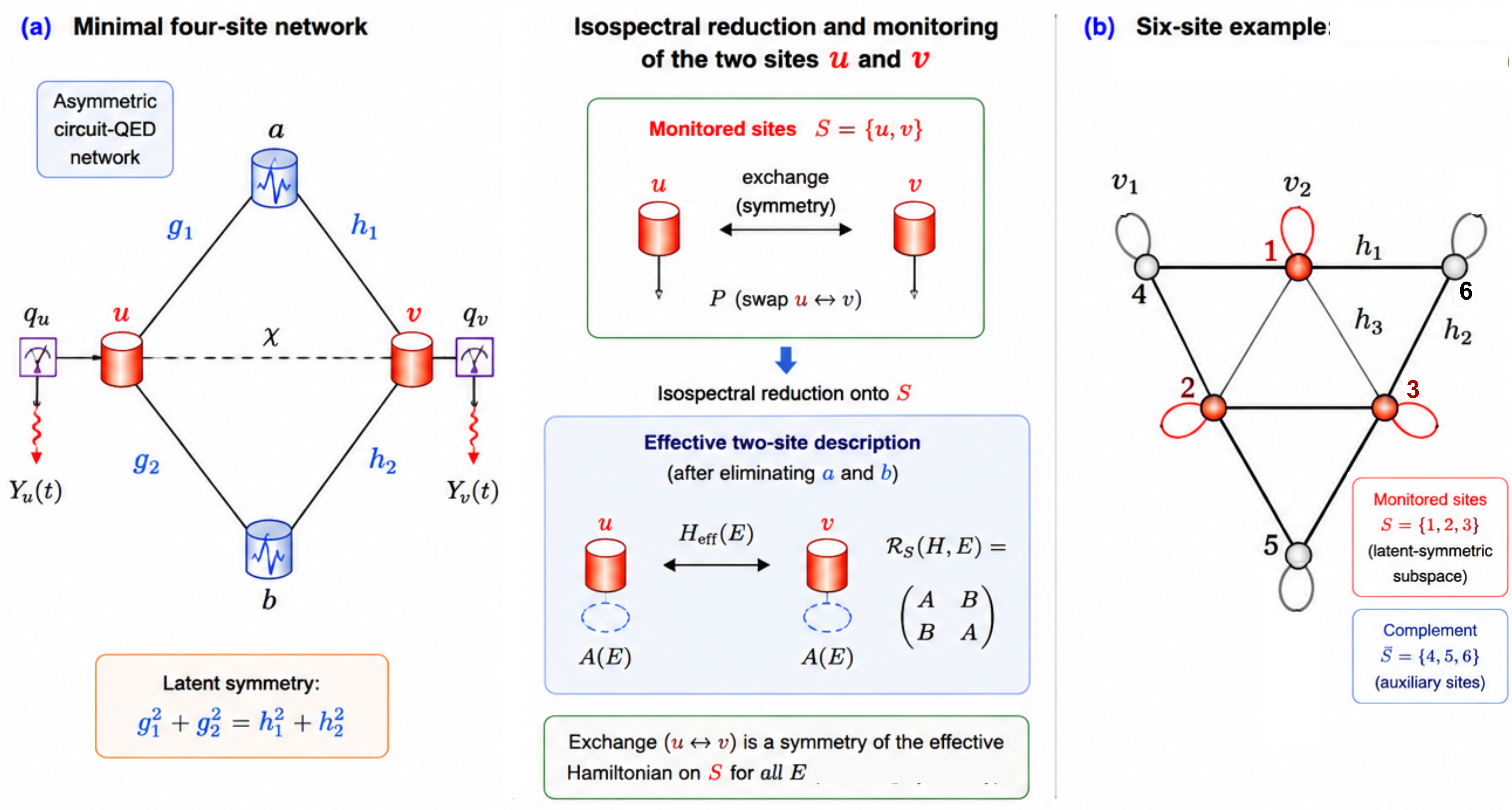}
\caption{
Latent symmetry and monitored-network models.
(a) Minimal four-resonator circuit-QED network considered in the main
text. Two monitored resonators $u$ and $v$ are coupled through the
auxiliary resonators $a$ and $b$ with generally different coupling
patterns $(g_1,g_2)$ and $(h_1,h_2)$. The condition
$g_1^2+g_2^2=h_1^2+h_2^2$ produces an exact latent exchange symmetry
between $u$ and $v$, despite the absence of a corresponding geometrical
symmetry of the physical network. Continuous QND monitoring of the
occupations of $u$ and $v$ yields the measurement records $Y_u(t)$
and $Y_v(t)$.
Center: isospectral reduction onto the monitored subspace
$S=\{u,v\}$ eliminates the auxiliary sites and yields an effective
two-site Hamiltonian with equal self-energies $A(E)$ and coupling
$B(E)$, which is invariant under exchange $u\leftrightarrow v$ for
all $E$.
(b) Six-site latent-symmetric network adapted from
Ref.~\cite{LS5}. The three distinguished sites
$S=\{1,2,3\}$ are shown in red, while the complementary sites are
shown in gray. Although the full six-site network possesses only a
$C_3$ geometrical symmetry for generic couplings $h_1 \neq h_2$, isospectral
reduction onto $S$ yields an effective Hamiltonian with the larger
non-Abelian $D_3$ permutation symmetry. This example illustrates that
latent symmetry can involve higher-dimensional distinguished
subspaces and is not restricted to the pairwise exchange symmetry of
the minimal four-site model.
}
\label{fig:network}
\end{figure*}

Throughout this work we restrict ourselves to the single-excitation
sector, $N=1$, corresponding experimentally to a single microwave
photon in the network. This is the minimal setting in which the
interplay between latent symmetry and quantum Zeno dynamics can be
exposed without additional many-body effects. The construction extends
naturally to higher-excitation sectors, where interactions and
many-body measurement effects provide interesting directions for
future study.\\		
Introducing the basis states
\begin{equation}
\ket{j}=\hat a_j^\dagger\ket{0},
\qquad
j=u,v,a,b,
\label{eq:singlephotonbasis}
\end{equation}
where $\ket{0}$ denotes the vacuum, the Hamiltonian in the basis
$(\ket{u},\ket{v},\ket{a},\ket{b})$ becomes
\begin{equation}
H_0=
\begin{pmatrix}
\epsilon_0&0&g_1&g_2\\
0&\epsilon_0&h_1&h_2\\
g_1&h_1&\Delta&0\\
g_2&h_2&0&\Delta
\end{pmatrix}.
\label{eq:Hcell}
\end{equation}
The four coupling amplitudes are generally different. We impose only
the latent-symmetry condition
\begin{equation}
g_1^2+g_2^2=h_1^2+h_2^2\equiv G^2,
\label{eq:latent_condition}
\end{equation}
where $G$ defines the overall coupling scale. A convenient
parametrization of the two coupling vectors is
\begin{equation}
\bm g
=
G
\begin{pmatrix}
\cos\alpha\\
\sin\alpha
\end{pmatrix},
\qquad
\bm h
=
G
\begin{pmatrix}
\cos\beta\\
\sin\beta
\end{pmatrix},
\label{eq:angular_parameterization}
\end{equation}
where
\begin{equation}
\bm g=
\begin{pmatrix}
g_1\\
g_2
\end{pmatrix},
\qquad
\bm h=
\begin{pmatrix}
h_1\\
h_2
\end{pmatrix}.
\label{eq:ghvectors}
\end{equation}
Equation~\eqref{eq:angular_parameterization} automatically satisfies
Eq.~\eqref{eq:latent_condition}, while the angles $\alpha$ and
$\beta$ independently control the coupling patterns of the two
distinguished resonators.
For generic $\alpha$ and $\beta$, the two coupling patterns are not
related by a permutation of the auxiliary resonators. An ordinary
geometrical exchange that swaps $u$ and $v$ while leaving the auxiliary
set $\{a,b\}$ invariant would require either identical coupling vectors
or their permutation, corresponding respectively to
\begin{equation}
\bm h=\bm g
\end{equation}
or
\begin{equation}
\bm h=
\begin{pmatrix}
g_2\\
g_1
\end{pmatrix}.
\end{equation}
Away from these special cases, the physical environments of $u$ and
$v$ are geometrically inequivalent even though
Eq.~\eqref{eq:latent_condition} remains exactly satisfied.

Thus there exists no permutation matrix $P$ satisfying
simultaneously
\begin{equation}
[H_0,P]=0,
\qquad
P\ket{u}=\ket{v}.
\label{eq:noordinary}
\end{equation}
The physical circuit therefore possesses no ordinary geometrical
symmetry exchanging resonators $u$ and $v$.

The latent symmetry becomes manifest upon performing an exact
isospectral reduction of the Hamiltonian onto the distinguished
resonators $u$ and $v$. This reduction eliminates the auxiliary
resonators $a$ and $b$ from the reduced representation without
introducing any approximation to the full four-resonator dynamics.
Let
\begin{equation}
S=\{u,v\},
\qquad
\bar S=\{a,b\}.
\end{equation}
The exact isospectral reduction onto $S$ is
\begin{equation}
\mathcal R(E)
=
H_{SS}
+
H_{S\bar S}
(E-H_{\bar S\bar S})^{-1}
H_{\bar S S}.
\label{eq:schur}
\end{equation}
Since
\begin{equation}
H_{\bar S\bar S}=\Delta I,
\end{equation}
one obtains
\begin{equation}
\mathcal R(E)
=
\epsilon_0 I
+
\frac{1}{E-\Delta}
\begin{pmatrix}
g_1^2+g_2^2&
g_1h_1+g_2h_2\\
g_1h_1+g_2h_2&
h_1^2+h_2^2
\end{pmatrix}.
\label{eq:reduced_general}
\end{equation}
\begingroup\color{black}
It is worth noting that the equality of the bare resonance frequencies
of the two distinguished resonators is part of the exact latent-symmetry
condition for the present minimal model. Indeed, if their frequencies
are allowed to differ, $\epsilon_u\neq\epsilon_v$, the two diagonal
elements of the reduced Hamiltonian become
$\epsilon_u+\|\bm g\|^2/(E-\Delta)$ and
$\epsilon_v+\|\bm h\|^2/(E-\Delta)$, respectively. Their equality
for every $E$ requires separately $\epsilon_u=\epsilon_v$ and
$\|\bm g\|=\|\bm h\|$. Thus, unequal coupling strengths can
compensate a frequency mismatch at a selected energy, but cannot restore
the energy-independent latent exchange symmetry considered here.
\endgroup

Using Eq.~\eqref{eq:latent_condition}, the reduced Hamiltonian takes
the form
\begin{equation}
\mathcal R(E)=A(E)I+B(E)\tau^x,
\label{eq:reduced_symmetric}
\end{equation}
where
\begin{equation}
\tau^x
=
\ket{u}\bra{v}
+
\ket{v}\bra{u}
=
\begin{pmatrix}
0&1\\
1&0
\end{pmatrix}
\label{eq:taux}
\end{equation}
is the exchange operator in the reduced $(u,v)$ subspace, and
\begin{align}
A(E)
&=
\epsilon_0+
\frac{g_1^2+g_2^2}{E-\Delta},
\label{eq:A}\\
B(E)
&=
\frac{g_1h_1+g_2h_2}{E-\Delta}.
\label{eq:B}
\end{align}
Consequently,
\begin{equation}
[\mathcal R(E),\tau^x]=0
\label{eq:latent_commutator}
\end{equation}
for every $E$. Thus the exchange $u\leftrightarrow v$ is an exact
symmetry of the reduced problem even though it is not a geometrical
symmetry of the physical four-resonator circuit. This result does not
involve any assumption on the magnitude of
$\Delta-\epsilon_0$ and, in particular, does not require a dispersive
regime between the resonator modes.

\subsection{Hidden symmetry operator}

The latent symmetry revealed by the reduced Hamiltonian can also be
represented directly in the complete single-excitation Hilbert space.
Using the coupling vectors defined in Eq.~\eqref{eq:ghvectors},
condition~\eqref{eq:latent_condition} reads
\begin{equation}
\|\bm g\|=\|\bm h\|
\equiv G.
\label{eq:equalnorm}
\end{equation}
Since the two vectors have equal norm, there exists an orthogonal
reflection $R$ acting in the auxiliary subspace such that
\begin{equation}
R\bm g=\bm h,
\qquad
R\bm h=\bm g,
\qquad
R^2=I.
\label{eq:Rproperties}
\end{equation}
For $\bm g\neq\bm h$, one explicit choice is the Householder
reflection
\begin{equation}
R
=
I-
2\frac{(\bm g-\bm h)(\bm g-\bm h)^T}
{\|\bm g-\bm h\|^2}.
\label{eq:Rexplicit}
\end{equation}
The corresponding hidden symmetry operator on the complete
single-excitation Hilbert space is
\begin{equation}
Q=
\begin{pmatrix}
\tau^x&0\\
0&R
\end{pmatrix},
\label{eq:Qfull}
\end{equation}
which satisfies
\begin{equation}
Q^2=I,
\qquad
[H_0,Q]=0.
\label{eq:Qsymmetry}
\end{equation}
Its action on the distinguished resonators is
\begin{equation}
Q\ket{u}=\ket{v},
\qquad
Q\ket{v}=\ket{u},
\label{eq:Quv}
\end{equation}
whereas its action on the auxiliary resonators generally transforms
$\ket{a}$ and $\ket{b}$ into linear combinations of one another.
Hence $Q$ is not, in general, a permutation of the physical
resonators: it realizes the exchange of $u$ and $v$ through a hidden
transformation of the auxiliary subspace.
This full-space representation will be useful in the next sections because the same
operator $Q$ also constrains the monitored dynamics and the resulting
quantum-Zeno subspaces. 

\section{Continuous QND monitoring}

We now consider continuous quantum-nondemolition (QND) monitoring of
the microwave-photon occupations of resonators $u$ and $v$. Here QND
refers to the measurement process itself: the readout acquires
information about the local photon occupations without intentionally
absorbing the excitation, although these occupations remain dynamical
under the coherent inter-resonator coupling. A natural circuit-QED implementation
is obtained by dispersively coupling each distinguished resonator to an
ancillary superconducting qubit (see Fig.~\ref{fig:network}). Denoting
the two ancillas by $q_u$ and $q_v$, the corresponding dispersive
interactions have the generic form
\begin{equation}
\hat H_{\rm disp}
=
\chi_u\hat n_u\hat\sigma_z^{(u)}
+
\chi_v\hat n_v\hat\sigma_z^{(v)},
\label{eq:Hdisp}
\end{equation}
where $\hat\sigma_z^{(\mu)}$ is the Pauli-$z$ operator of the ancillary
qubit $q_\mu$ ($\mu=u,v$), and $\chi_\mu$ denotes the corresponding
dispersive coupling strength. The state-dependent frequency shift of each ancilla, and hence its
continuously measured readout signal, carries information about the
photon occupation of the associated resonator.
The ancillas are continuously interrogated through conventional
circuit-QED readout channels. In the regime in which the ancilla and
readout dynamics can be eliminated, the resulting effective
description is a continuous QND measurement of $\hat n_u$ and
$\hat n_v$. We take the two effective measurement strengths to be
equal and denote them by $\gamma$. The corresponding measurement
operators are
\begin{equation}
\hat L_u=\sqrt{\gamma}\,\hat n_u,
\qquad
\hat L_v=\sqrt{\gamma}\,\hat n_v.
\label{eq:jumps}
\end{equation}
\begingroup\color{black}

The equality of the two measurement strengths is not merely a
simplifying assumption, but is required for the monitoring itself to
preserve the latent exchange symmetry. More generally, for strengths
$\gamma_u$ and $\gamma_v$ the measurement contribution to the
unconditional dynamics is
$\gamma_u\mathcal D[n_u]+\gamma_v\mathcal D[n_v]$, where
$\mathcal D[L]\rho=L\rho L^\dagger-
\{L^\dagger L,\rho\}/2$ denotes the standard Lindblad dissipator.
Since the hidden transformation exchanges $n_u$ and $n_v$, this
contribution is invariant under $Q$ only when
$\gamma_u=\gamma_v$. If only $u$ is monitored, the measurement instead
resolves the two subspaces $\mathrm{span}\{\ket{u}\}$ and
$\mathrm{span}\{\ket{v},\ket{a},\ket{b}\}$; in particular,
$\ket{v}$ no longer constitutes a separate measurement-selected Zeno
sector. The symmetry-enforced equality of the two polarized-sector
escape statistics derived below therefore does not apply to such an
asymmetric monitoring protocol.
\endgroup
Since
\begin{equation}
[\hat n_u,\hat N]=[\hat n_v,\hat N]=0,
\label{eq:measurement_number}
\end{equation}
the ideal monitoring induces dephasing but neither creates nor removes
microwave photons from the resonator network.
We stress that the dispersive approximation invoked here concerns only
the coupling between each monitored resonator and its readout ancilla.
It is unrelated to the detuning $\Delta-\epsilon_0$ between the four
network resonators. The resonator modes themselves need not be
dispersively separated, and all four remain explicitly involved in the
network dynamics.\\
When the measurement records are discarded, the unconditional density
matrix evolves according to the Lindblad master equation
\cite{QND1,QND2},
\begin{equation}
\frac{d\rho}{dt}
=
-i[H_0,\rho]
+
\gamma\mathcal D[n_u]\rho
+
\gamma\mathcal D[n_v]\rho \equiv \mathcal L \rho,
\label{eq:master_full}
\end{equation}
where
\begin{equation}
\mathcal D[L]\rho
=
L\rho L^\dagger
-\frac{1}{2}
\{L^\dagger L,\rho\}
\label{eq:dissipator}
\end{equation}
is the Lindblad dissipator and
\begin{equation}
n_u=\ket{u}\bra{u},
\qquad
n_v=\ket{v}\bra{v}
\label{eq:number_single}
\end{equation}
are the photon-number operators in the single-excitation sector.
Importantly, the monitoring preserves the latent symmetry introduced
in Sec.~II. Indeed,
\begin{equation}
Qn_uQ^{-1}=n_v,
\qquad
Qn_vQ^{-1}=n_u.
\label{eq:Qmeasure}
\end{equation}
Since the two measurement channels have equal strength and
$[H_0,Q]=0$, the complete unconditional dynamics is invariant under
the latent transformation. Defining
\begin{equation}
\mathcal Q(\rho)=Q\rho Q^{-1},
\end{equation}
the single-excitation Liouvillian satisfies
\begin{equation}
\mathcal Q\mathcal L\mathcal Q^{-1}
=
\mathcal L.
\label{eq:Liouvillian_symmetry}
\end{equation}
An initial density matrix is latent symmetric when
\begin{equation}
Q\rho(0)Q^{-1}=\rho(0).
\label{eq:latent_symmetric_state}
\end{equation}
For such an initial condition, the symmetry is preserved by the
unconditional evolution,
\begin{equation}
Q\rho(t)Q^{-1}=\rho(t)
\label{eq:latent_symmetry_preserved}
\end{equation}
at all later times. Since the occupation imbalance is odd under the
hidden exchange,
\begin{equation}
Q(n_u-n_v)Q^{-1}
=
-(n_u-n_v),
\label{eq:imbalance_odd}
\end{equation}
it follows that
\begin{equation}
\langle n_u-n_v\rangle_{\rm ens}=0.
\label{eq:ensemble_zero}
\end{equation}
Thus continuous monitoring does not distinguish between the two
geometrically inequivalent resonators at the ensemble level when the
initial state respects the latent symmetry.
It is also worth noting that the stationary state of the present
unconditional dynamics does not by itself reveal the latent symmetry.
Indeed, the Lindbladian is unital and, for the generic connected network considered here, the unique stationary state in the single-excitation sector is the maximally mixed state
 $\rho_{\rm ss}=I_4/4$. 
This state remains stationary even
when the latent-symmetry condition is broken. Latent symmetry is
therefore encoded not in the stationary populations, but in the
dynamical relations among relaxation modes and, most directly here,
in the statistics of individual measurement trajectories.

\section{Latent symmetry in quantum Zeno dynamics}

\subsection{Quantum trajectories and Zeno subspaces}

The Lindblad equation~\eqref{eq:master_full} describes the
\emph{unconditional} state obtained when the monitoring is performed
but its outcomes are discarded. To resolve the dynamics associated
with individual continuous-measurement records, we introduce the
conditional density matrix $\rho_c(t)$
\cite{QND1,QND2,QND3,QED4,QED8}. The unconditional state is recovered
as
\begin{equation}
\rho(t)=\overline{\rho_c(t)},
\label{eq:conditional_average}
\end{equation}
where the overbar denotes an average over all measurement records.\\
For unit-efficiency diffusive monitoring, the conditional
single-excitation dynamics is described by the stochastic master
equation \cite{QND1,QND2}
\begin{align}
d\rho_c={}&
-i[H_0,\rho_c]\,dt
+\gamma\mathcal D[n_u]\rho_c\,dt
+\gamma\mathcal D[n_v]\rho_c\,dt
\nonumber\\
&+
\sqrt{\gamma}\,
\mathcal H[n_u]\rho_c\,dW_u
+
\sqrt{\gamma}\,
\mathcal H[n_v]\rho_c\,dW_v ,
\label{eq:SME_full}
\end{align}
where
\begin{equation}
\mathcal H[L]\rho
=
L\rho+\rho L^\dagger
-
\mathrm{Tr}[(L+L^\dagger)\rho]\,\rho
\label{eq:innovation}
\end{equation}
is the measurement-innovation superoperator and $dW_u,dW_v$ are
independent real Wiener increments satisfying
\begin{equation}
\overline{dW_\mu}=0,
\qquad
dW_\mu dW_\nu=\delta_{\mu\nu}\,dt.
\label{eq:Wiener_uv}
\end{equation}
Equation~\eqref{eq:SME_full} is an effective stochastic description
after elimination of the readout ancillas; all four network resonators
remain explicitly included in $H_0$.\\
The corresponding continuous measurement records can be written as
\begin{align}
dY_u
&=
2\sqrt{\gamma}\,
\langle n_u\rangle_c\,dt+dW_u,
\label{eq:Yu}\\
dY_v
&=
2\sqrt{\gamma}\,
\langle n_v\rangle_c\,dt+dW_v.
\label{eq:Yv}
\end{align}
The two records jointly resolve the monitored occupations
$(n_u,n_v)$, while their difference directly probes the occupation
imbalance. Introducing
\begin{equation}
dY_-=\frac{dY_u-dY_v}{\sqrt{2}},
\qquad
dW_-=\frac{dW_u-dW_v}{\sqrt{2}},
\end{equation}
one obtains
\begin{equation}
dY_-
=
\sqrt{2\gamma}\,m_c(t)\,dt+dW_-,
\label{eq:measurement_record}
\end{equation}
where
\begin{equation}
m_c(t)
=
\mathrm{Tr}
\left[
\rho_c(t)(n_u-n_v)
\right]
\label{eq:mc}
\end{equation}
is the conditional latent polarization.
The quantity $m_c$ ranges from $-1$ to $1$. The limiting values
$m_c=+1$ and $m_c=-1$ correspond to a microwave photon localized on
resonator $u$ and $v$, respectively. A state entirely confined to the
auxiliary $(a,b)$ subspace has $m_c=0$, although $m_c=0$ alone does
not uniquely identify that subspace for a generic conditional state:
it can also result from equal nonzero occupations of $u$ and $v$.
The two separate records $Y_u$ and $Y_v$, however, distinguish these
situations and resolve the auxiliary measurement outcome
$(n_u,n_v)=(0,0)$. In the strong-monitoring regime, where the
conditional dynamics becomes localized near the measurement
eigenspaces, plateaus near $m_c=0$ can therefore be associated with
the auxiliary Zeno sector.\\		
The conditional dynamics inherits the latent symmetry in a statistical
sense. Using Eq.~\eqref{eq:Qmeasure}, one finds
\begin{align}
\mathcal D[n_u](Q\rho Q^{-1})
&=
Q\mathcal D[n_v](\rho)Q^{-1},
\label{eq:Dcovariance}\\
\mathcal H[n_u](Q\rho Q^{-1})
&=
Q\mathcal H[n_v](\rho)Q^{-1},
\label{eq:Hcovariance}
\end{align}
together with the corresponding relations under
$u\leftrightarrow v$. Since $[H_0,Q]=0$, the stochastic master
equation~\eqref{eq:SME_full} is therefore covariant under the combined
transformation
\begin{equation}
\rho_c\longrightarrow Q\rho_cQ^{-1},
\qquad
dW_u\longleftrightarrow dW_v.
\label{eq:trajectory_symmetry}
\end{equation}
More precisely, if $\rho_c(t)$ is a conditional trajectory generated
by a given pair of Wiener histories $(dW_u,dW_v)$, then
$Q\rho_c(t)Q^{-1}$ is the latent-related trajectory generated by the
exchanged histories $(dW_v,dW_u)$.\\
Because $dW_u$ and $dW_v$ are independent and statistically
identical, the original and exchanged noise histories occur with the
same probability. Together with the odd transformation of the
occupation imbalance, Eq.~\eqref{eq:imbalance_odd}, this implies that,
for a latent-symmetric initial state, every trajectory with
polarization history $m_c(t)$ has an equally probable latent-related
trajectory with history $-m_c(t)$. Denoting by
$\mathcal P[m_c(\cdot)]$ the probability functional of a complete
polarization history, one has
\begin{equation}
\mathcal P[m_c(\cdot)]
=
\mathcal P[-m_c(\cdot)].
\label{eq:path_symmetry}
\end{equation}
Averaging over trajectories therefore reproduces the vanishing
ensemble imbalance of Eq.~\eqref{eq:ensemble_zero}. This statistical
pairing of conditional trajectories is exact for
arbitrary $\Delta$, including the resonant regime
$\Delta=\epsilon_0$.\\
We emphasize that resolving individual quantum trajectories is not
necessary merely to establish the presence of latent symmetry. Since
the unconditional Liouvillian $\mathcal L$ commutes with the latent transformation,
initial states related by $Q$ exhibit correspondingly related
ensemble-averaged relaxation dynamics. Trajectory resolution provides
additional information: it makes the measurement-selected dynamical
sectors individually accessible and allows their residence, escape,
and switching statistics to be resolved. As shown below, in the
strong-monitoring regime these sectors become quantum-Zeno subspaces,
and latent symmetry manifests itself as an exact statistical pairing
of geometrically inequivalent Zeno sectors.\\ 
Quantum Zeno dynamics is naturally formulated in terms of the
eigenspaces of the measured observables
\cite{QZ3,QZ5,QZ8,QZ10,QZ15}. Since $n_u$ and $n_v$ commute, their
three distinct joint eigenspaces in the single-excitation Hilbert
space are
\begin{align}
\mathcal Z_u
&=
\mathrm{span}\{\ket{u}\},
\label{eq:Zu}\\
\mathcal Z_v
&=
\mathrm{span}\{\ket{v}\},
\label{eq:Zv}\\
\mathcal Z_0
&=
\mathrm{span}\{\ket{a},\ket{b}\}.
\label{eq:Z0}
\end{align}
They correspond respectively to the joint measurement outcomes
\begin{equation}
(n_u,n_v)
=
(1,0),\quad(0,1),\quad(0,0).
\label{eq:joint_measurement_outcomes}
\end{equation}

\begin{figure*}[t]
\centering
\includegraphics[width=\textwidth]{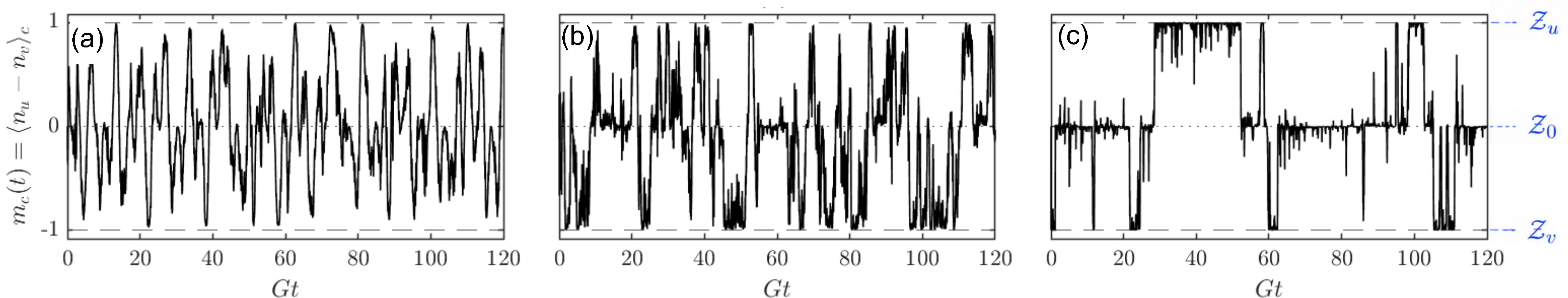}
\caption{
Emergence of quantum-Zeno subspaces in individual continuously
monitored trajectories. Conditional latent polarization
$m_c(t)=\langle n_u-n_v\rangle_c$ for
(a) $\gamma/G=0.5$, (b) $\gamma/G=3$, and
(c) $\gamma/G=15$. The coupling vectors are parametrized according to
Eq.~\eqref{eq:angular_parameterization} with
$\alpha=10^\circ$ and $\beta=55^\circ$, and the resonators are taken
at resonance, $\Delta=\epsilon_0$. The initial state is the
latent-even state
$\ket{\psi_c(0)}=(\ket{u}+\ket{v})/\sqrt{2}$.
Trajectories are obtained by Euler--Maruyama integration of the
diffusive stochastic Schr\"odinger equation~\eqref{eq:SSE_numerical}
over $0\leq Gt\leq120$, with time step
$G\,dt=2\times10^{-4}$. Increasing the monitoring strength
progressively resolves the three joint measurement eigenspaces. In
the strong-monitoring regime, long plateaus near
$m_c\simeq+1$, $0$, and $-1$ correspond, respectively, to
$\mathcal Z_u$, $\mathcal Z_0$, and $\mathcal Z_v$; identification
of $\mathcal Z_0$ is supplied by the joint record $(Y_u,Y_v)$.
}
\label{fig:trajectories}
\end{figure*}
Thus the two polarized sectors are distinguished by the differential
record, whereas the full pair of measurement records also identifies
the two-dimensional auxiliary sector.\\
Figure~\ref{fig:trajectories} illustrates the emergence of these Zeno
subspaces in individual quantum trajectories as the monitoring
strength is increased. The trajectories are obtained by numerically
unraveling the stochastic master equation~\eqref{eq:SME_full} in the
single-excitation sector. Since the monitoring efficiency is taken to
be unity and the initial state is pure, the conditional state remains
pure and can be represented as
$\rho_c(t)=\ket{\psi_c(t)}\bra{\psi_c(t)}$. We therefore integrate the
equivalent normalized diffusive stochastic Schr\"odinger equation
\begin{align}
d\ket{\psi_c}
={}&
\left[
-iH_0
-\frac{\gamma}{2}
\sum_{\mu=u,v}
\left(
n_\mu-\langle n_\mu\rangle_c
\right)^2
\right]
\ket{\psi_c}\,dt
\nonumber\\
&+
\sqrt{\gamma}
\sum_{\mu=u,v}
\left(
n_\mu-\langle n_\mu\rangle_c
\right)
\ket{\psi_c}\,dW_\mu ,
\label{eq:SSE_numerical}
\end{align}
where
\begin{equation}
\langle n_\mu\rangle_c
=
\bra{\psi_c}n_\mu\ket{\psi_c}.
\end{equation}
Equation~\eqref{eq:SSE_numerical} is integrated using the
Euler--Maruyama method with independent Gaussian Wiener increments
$dW_\mu=\sqrt{dt}\,\xi_\mu$, where $\xi_\mu$ are independent
zero-mean, unit-variance normal random variables. The state is
renormalized after each integration step to suppress accumulated
numerical errors. No reduction or adiabatic elimination of the
auxiliary resonators is performed in the simulations: the complete
four-dimensional single-excitation Hamiltonian~\eqref{eq:Hcell} is
used throughout.\\
For weak monitoring [Fig.~\ref{fig:trajectories}(a)], the measurement
only weakly resolves the occupation sectors and the conditional
polarization $m_c(t)$ fluctuates broadly. At intermediate strength
[Fig.~\ref{fig:trajectories}(b)], increasingly pronounced intervals
near $m_c=\pm1$ and $m_c=0$ emerge. In the strong-monitoring regime
[Fig.~\ref{fig:trajectories}(c)], the measurement records resolve the
three sectors $\mathcal Z_u$, $\mathcal Z_0$, and $\mathcal Z_v$,
and the polarization displays long plateaus near $+1$, $0$, and $-1$
interrupted by relatively rapid transitions. This progressive
dynamical separation of the joint measurement eigenspaces is the
trajectory-level manifestation of quantum Zeno dynamics.

\subsection{Latent-symmetry-constrained Zeno dynamics}

To characterize the Zeno dynamics constrained by the latent symmetry, we introduce the projectors
\begin{equation}
\Pi_u=n_u,\qquad
\Pi_v=n_v,\qquad
\Pi_0=I-n_u-n_v,
\label{eq:Zeno_projectors}
\end{equation}
such that
\begin{equation}
\Pi_u+\Pi_v+\Pi_0=I.
\end{equation}
In the ideal quantum-Zeno limit, coherences between different
measurement eigenspaces are eliminated and transitions between them
are suppressed. The leading coherent evolution is generated by the
projected Zeno Hamiltonian
\begin{equation}
H_{\rm Z}
=
\sum_{\alpha=u,v,0}
\Pi_\alpha H_0\Pi_\alpha,
\label{eq:Zeno_Hamiltonian}
\end{equation}
which for the present model becomes
\begin{equation}
H_{\rm Z}
=
\epsilon_0(\Pi_u+\Pi_v)
+
\Delta\Pi_0.
\label{eq:Zeno_Hamiltonian_explicit}
\end{equation}
Thus the three measurement subspaces become dynamically disconnected
in the ideal Zeno limit, whereas at large but finite $\gamma$ rare
Hamiltonian-induced transitions remain possible
\cite{QZ5,QZ8,QZ9,QZ14,QZ15}.
The central result follows from the action of the latent symmetry on
the Zeno decomposition. Using Eq.~\eqref{eq:Qmeasure}, the Zeno
projectors transform as
\begin{equation}
Q\Pi_uQ^{-1}=\Pi_v,
\qquad
Q\Pi_vQ^{-1}=\Pi_u,
\qquad
Q\Pi_0Q^{-1}=\Pi_0.
\label{eq:Q_Zeno_projectors}
\end{equation}
Since the Zeno subspaces are the ranges of the corresponding
projectors,
\begin{equation}
\mathcal Z_\alpha=\operatorname{Ran}\Pi_\alpha,
\qquad
\alpha=u,v,0,
\end{equation}
Eq.~\eqref{eq:Q_Zeno_projectors} directly determines how the hidden
symmetry acts on the subspaces themselves. For example, if
$\ket{\psi}\in\mathcal Z_u$, then
$\Pi_u\ket{\psi}=\ket{\psi}$ and, using
$\Pi_vQ=Q\Pi_u$,
\begin{equation}
\Pi_v Q\ket{\psi}
=
Q\Pi_u\ket{\psi}
=
Q\ket{\psi}.
\end{equation}
Thus $Q\ket{\psi}\in\mathcal Z_v$, and hence
$Q\mathcal Z_u=\mathcal Z_v$. Analogously,
$Q\mathcal Z_v=\mathcal Z_u$, whereas the auxiliary subspace is
mapped onto itself. We can therefore summarize the action of the
latent symmetry as
\begin{equation}
Q\mathcal Z_u=\mathcal Z_v,
\qquad
Q\mathcal Z_v=\mathcal Z_u,
\qquad
Q\mathcal Z_0=\mathcal Z_0.
\label{eq:Q_Zeno_spaces}
\end{equation}
Measurement thus creates the Zeno subspaces, while latent symmetry
organizes them: the two one-dimensional sectors $\mathcal Z_u$ and
$\mathcal Z_v$ form an exact symmetry-related pair, whereas the
two-dimensional auxiliary sector is invariant as a whole.
This structure is fundamentally different from that generated by an
ordinary geometrical symmetry. A geometrical exchange of $u$ and $v$
would be implemented by a permutation of the complete physical
resonator basis. Here, instead, $Q$ exchanges the physical states
$\ket{u}$ and $\ket{v}$ while acting as a nontrivial rotation in the
auxiliary $(a,b)$ subspace. Consequently, the individual physical
resonators $a$ and $b$ need not be symmetry related even though
$\mathcal Z_u$ and $\mathcal Z_v$ are exactly exchanged. Quantum Zeno
dynamics therefore provides a direct operational manifestation of a
symmetry that is hidden from the physical geometry of the network.

\subsection{Zeno escape and residence statistics}

The latent relation between $\mathcal Z_u$ and $\mathcal Z_v$ has a
direct quantitative manifestation in their escape statistics. The path
symmetry established above already implies identical first-passage
statistics for the two sectors when the initial conditions and exit
criteria are related by $Q$. In the Zeno regime, their common escape
rate can be obtained explicitly. Let
\begin{equation}
\delta_\Delta=\Delta-\epsilon_0
\end{equation}
denote the detuning between the distinguished and auxiliary
resonators. Under the Lindblad convention of
Eq.~\eqref{eq:master_full}, coherences connecting $\ket{u}$ or
$\ket{v}$ to the auxiliary subspace decay at the
measurement-induced rate $\gamma/2$.

In the Zeno regime, these rapidly decaying intersector coherences can
be adiabatically eliminated. For resonator $u$, the two coherent
couplings $g_1$ and $g_2$ to the auxiliary subspace then generate
incoherent transitions whose total rate is
\begin{equation}
\Gamma_u
\simeq
\frac{
\gamma(g_1^2+g_2^2)
}{
\delta_\Delta^2+(\gamma/2)^2
}.
\label{eq:Gamma_u}
\end{equation}
Likewise, the escape rate from $\mathcal Z_v$ is
\begin{equation}
\Gamma_v
\simeq
\frac{
\gamma(h_1^2+h_2^2)
}{
\delta_\Delta^2+(\gamma/2)^2
}.
\label{eq:Gamma_v}
\end{equation}
Using the latent-symmetry condition~\eqref{eq:latent_condition}, the
adiabatic result is consistent with the exact trajectory pairing and
gives
\begin{equation}
\Gamma_u=\Gamma_v.
\label{eq:equal_escape}
\end{equation}
Thus latent symmetry enforces identical Zeno escape statistics for two
geometrically inequivalent physical sectors despite their different
microscopic coupling patterns.

In the strong-monitoring limit,
\begin{equation}
\gamma\gg|\delta_\Delta|,|g_j|,|h_j|,
\end{equation}
the common escape rate reduces to
\begin{equation}
\Gamma_u=\Gamma_v
\simeq
\frac{4G^2}{\gamma},
\label{eq:Gamma_Zeno}
\end{equation}
and the corresponding mean residence times satisfy
\begin{equation}
\tau_u=\tau_v
\simeq
\frac{\gamma}{4G^2}.
\label{eq:residence}
\end{equation}
The quantum Zeno effect therefore suppresses escape as $1/\gamma$,
while latent symmetry fixes the equality of the characteristic time
scales of the two polarized sectors.

\begin{figure}[t]
\centering
\includegraphics[width=\columnwidth]{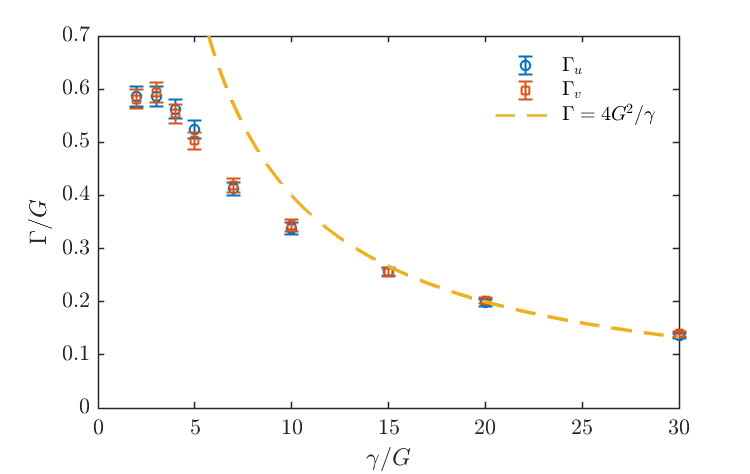}
\caption{
Latent-symmetry-constrained Zeno escape rates. Numerically extracted
escape rates $\Gamma_u$ and $\Gamma_v$ from the polarized sectors
$\mathcal Z_u$ and $\mathcal Z_v$ as a function of the monitoring
strength $\gamma/G$. For each value of $\gamma$, the two rates are
independently estimated from $10^3$ first-passage trajectories
initialized in $\ket{u}$ and $\ket{v}$, respectively. First passage to
either of the other Zeno sectors is identified using the localization
threshold $p_{\rm th}=0.9$, with right-censored trajectories included
in the rate estimator. The coupling vectors are parametrized according
to Eq.~\eqref{eq:angular_parameterization} with
$\alpha=10^\circ$, $\beta=55^\circ$, and
$\Delta=\epsilon_0$. Numerical integration uses
$G\,dt=2\times10^{-4}$ and a maximum observation time
$GT_{\max}=60$. The independently extracted rates coincide within
statistical uncertainty and, in the strong-monitoring regime, approach
the Zeno asymptote $\Gamma_{\rm Z}=4G^2/\gamma$ (dashed line).
Error bars indicate the estimated standard errors.
}
\label{fig:zenoscaling}
\end{figure}

Figure~\ref{fig:zenoscaling} provides a quantitative test of these
predictions. The escape rates are extracted directly from first-passage
statistics of the stochastic dynamics. 
For each value of $\gamma$,
independent ensembles of $10^3$ trajectories are initialized in either
$\ket{u}$ or $\ket{v}$ and evolved according to the stochastic
Schr\"odinger equation~\eqref{eq:SSE_numerical}. Starting from
$\ket{u}$, the first exit from $\mathcal Z_u$ is registered when the
conditional state becomes localized in either of the other two Zeno
sectors, identified by $p_v>p_{\rm th}$ or
$p_0>p_{\rm th}$, where
$p_0=1-p_u-p_v$ and $p_{\rm th}=0.9$; the analogous criterion is used
for trajectories initialized in $\ket{v}$. Trajectories that do not
escape within the simulation window are treated as right-censored
realizations. We use throughout the exponential-rate estimator given by
the number of observed exits divided by the total accumulated exposure
time, including that of censored trajectories. In the strong-monitoring
regime, where the intersector dynamics becomes effectively Markovian,
this estimator directly yields the Zeno escape rate.

As shown in Fig.~\ref{fig:zenoscaling}, the independently extracted
rates $\Gamma_u$ and $\Gamma_v$ coincide within statistical
uncertainty. In the strong-monitoring regime, both approach the
asymptotic Zeno law of Eq.~\eqref{eq:Gamma_Zeno}. Deviations at smaller
$\gamma/G$ are expected because the adiabatic elimination of
intersector coherences is no longer justified outside the Zeno regime.
The numerical results thus demonstrate both the symmetry-enforced
pairing of the escape rates and their asymptotic $1/\gamma$
suppression.

A more stringent test is obtained by changing the microscopic
coupling geometry while preserving the latent symmetry. Within the
angular parametrization of Eq.~\eqref{eq:angular_parameterization},
the angles $\alpha$ and $\beta$ can be varied while the equal-norm
condition $\|\bm g\|=\|\bm h\|=G$ remains exactly satisfied.
Consequently, the local environments of $u$ and $v$ can be
continuously deformed without destroying the latent relation between
their Zeno sectors.

\begin{figure}[t]
\centering
\includegraphics[width=\columnwidth]{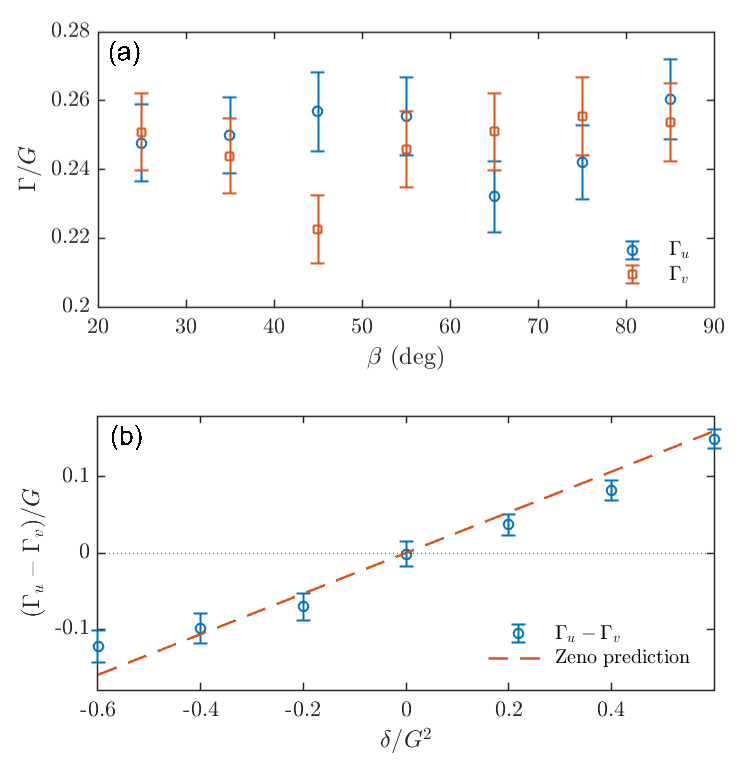}
\caption{
Control of Zeno escape dynamics by latent symmetry.
(a) Escape rates $\Gamma_u$ and $\Gamma_v$ under deformation of the
microscopic coupling geometry along the latent-symmetric manifold.
The angle $\alpha=10^\circ$ is fixed while $\beta$ is varied from
$25^\circ$ to $85^\circ$ in steps of $10^\circ$, with
$\|\bm g\|=\|\bm h\|=G$ maintained throughout. Despite the resulting
geometrical inequivalence of the local coupling environments, the two
independently extracted rates remain equal within statistical
uncertainty.
(b) Escape-rate splitting upon breaking the latent symmetry. The
coupling directions are fixed at $\alpha=10^\circ$ and
$\beta=55^\circ$, while the norm of $\bm g$ is kept fixed,
$\|\bm g\|=G$, and that of $\bm h$ is varied according to
$\|\bm h\|=rG$. Consequently,
$\delta/G^2=1-r^2$, where
$\delta=g_1^2+g_2^2-h_1^2-h_2^2$, is varied from $-0.6$ to $0.6$
in steps of $0.2$. The splitting vanishes at the latent-symmetric
point $\delta=0$ and follows the Zeno prediction of
Eq.~\eqref{eq:rate_splitting} (dashed line).
In both panels $\gamma/G=15$, $\Delta=\epsilon_0$,
$p_{\rm th}=0.9$, and each rate is independently estimated from
$500$ first-passage trajectories. Numerical integration uses
$G\,dt=2\times10^{-4}$ and $GT_{\max}=60$. Error bars indicate the
estimated standard errors.
}
\label{fig:latentcontrol}
\end{figure}

This behavior is demonstrated in Fig.~\ref{fig:latentcontrol}(a),
where $\alpha$ is fixed and $\beta$ is varied over a broad range.
Although the microscopic coupling geometry changes substantially,
the independently determined escape rates remain equal within
statistical uncertainty. The equality is therefore not tied to a
particular geometrical configuration, but follows from the latent
symmetry.
Conversely, the latent symmetry can be deliberately broken by
violating the equal-norm condition. Introducing
\begin{equation}
\delta
=
g_1^2+g_2^2-h_1^2-h_2^2
=
\|\bm g\|^2-\|\bm h\|^2,
\label{eq:breaking}
\end{equation}
the Zeno-rate splitting is predicted to be
\begin{equation}
\Gamma_u-\Gamma_v
\simeq
\frac{\gamma\delta}
{\delta_\Delta^2+(\gamma/2)^2}.
\label{eq:rate_splitting}
\end{equation}
In Fig.~\ref{fig:latentcontrol}(b), this symmetry breaking is
implemented by keeping the coupling directions $\alpha$ and $\beta$
fixed, while setting $\|\bm g\|=G$ and
$\|\bm h\|=rG$. Thus $\delta/G^2=1-r^2$, with $r=1$
corresponding to the latent-symmetric point. The numerically extracted
rate splitting vanishes at this point and develops with either sign
as the relative coupling strength is varied, in good agreement with
Eq.~\eqref{eq:rate_splitting}. Together, the two panels
distinguish geometrical deformation that preserves the latent constraint
from genuine latent-symmetry breaking.

\begingroup\color{black}
An operational advantage of latent over ordinary geometrical symmetry
is the additional freedom that it leaves in the microscopic
architecture. A geometrical exchange of $u$ and $v$ would require their
coupling environments to be related by a physical permutation, whereas
latent symmetry constrains only the effective properties relevant to
the hidden symmetry. Consequently, the microscopic couplings can be
varied over a continuous family of geometrically inequivalent
configurations while the exact relation between the two Zeno sectors is
retained, as demonstrated in Fig.~4(a). More generally, in complex
networks a symmetry can emerge only after isospectral reduction over a
nontrivial internal structure, allowing selected sites or subspaces to
obey exact symmetry relations even when their microscopic environments
are not related by any geometrical transformation
\cite{LS18}. Latent symmetry may therefore provide additional
architectural flexibility when geometrical symmetry is incompatible
with connectivity, fabrication, spatial, or independent-control
requirements of a physical device
\endgroup

The predicted dynamics should be accessible with current circuit-QED
technology. In the Zeno regime the differential continuous readout
$Y_u(t)-Y_v(t)$ distinguishes the three plateau values associated with
the polarized and auxiliary sectors, while the full pair $(Y_u,Y_v)$
provides their unambiguous joint readout. The
symmetry-specific signature is the equality of the escape statistics
of $\mathcal Z_u$ and $\mathcal Z_v$ despite their inequivalent local
environments; deliberately breaking the equal-norm condition provides
a natural control measurement. Coherent inter-resonator couplings in
the sub-MHz to MHz range can coexist with MHz-scale dispersive
monitoring and microwave-resonator lifetimes of tens to hundreds of
microseconds or longer \cite{QED5,QED6,QED7}. For example, taking
$G/(2\pi)\simeq0.45~{\rm MHz}$ and
$\gamma/(2\pi)\simeq5~{\rm MHz}$ gives from
Eq.~\eqref{eq:residence} a characteristic Zeno residence time of order
$1~\mu{\rm s}$, leaving a substantial time window for resolving the
latent-symmetry-constrained Zeno dynamics before intrinsic photon loss
becomes important.
\begingroup\color{black}
It should be mentioned that, even whe intrinsic photon loss are included in the model, loss of excitation does not by itself destroy the
latent symmetry. In the presence of resonator decay, the vacuum state
must be included in the analysis and the unconditional master equation acquires the
additional contribution
$\sum_j\kappa_j\mathcal D[a_j]\rho$, where $kappa_j$ are the photon decay rates. The hidden transformation can be
extended to the zero- and one-excitation Hilbert space as
$\widetilde Q=|0\rangle\langle0|\oplus Q$, leaving the vacuum
invariant. Latent symmetry of the open dynamics then means covariance
of the complete Liouvillian under
$\widetilde{\mathcal Q}(\rho)=\widetilde Q\rho\widetilde Q^{-1}$.
Although photon loss breaks excitation-number conservation, this
Liouvillian symmetry is preserved when the loss channels themselves
transform covariantly under $\widetilde Q$. A simple sufficient
condition for the present network is
$\kappa_u=\kappa_v$ and $\kappa_a=\kappa_b$. In this case photon loss
introduces a symmetry-related decay from the two polarized Zeno
sectors to the vacuum without lifting their latent pairing. The
long-time stationary state is then the vacuum rather than the
maximally mixed state of the lossless single-excitation dynamics.
Conversely, unequal losses $\kappa_u\neq\kappa_v$ explicitly break
the covariance of the full Liouvillian and distinguish the two
polarized sectors. If decay to the vacuum is counted as an additional
exit event, the total observed escape rate in the Zeno regime is, to
leading order,
$\Gamma_\mu^{\rm tot}\simeq
\Gamma_\mu^{\rm Zeno}+\kappa_\mu$.
Thus unequal photon losses provide an additional experimentally
accessible symmetry-breaking control, distinct from the
Hamiltonian-induced inter-Zeno-sector transitions considered in
Figs.~3 and 4.

\endgroup

\section{Conclusions and discussion}
In conclusion, we have established a connection between latent symmetry
and quantum Zeno dynamics, showing that hidden symmetries of a physical
network can impose exact constraints on measurement-induced Zeno
subspaces. We have illustrated this interplay using a minimal
four-resonator network naturally realizable in circuit QED. Continuous
QND monitoring generates Zeno sectors associated with geometrically
inequivalent resonators that are exactly exchanged by the latent
symmetry and consequently display identical escape and residence
statistics. This equivalence persists under geometrical deformations
that preserve the latent symmetry and is lifted when the symmetry is
broken. The distinctive signature is therefore not Zeno localization
itself, but the symmetry-enforced relation between Zeno subspaces that
are inequivalent in the physical geometry.

The four-resonator network considered here is the minimal realization
of the mechanism, but the underlying principle is not restricted to
this geometry or to pairwise latent symmetries. As an example,
Fig.~\ref{fig:network}(b) shows the six-site network introduced in
Ref.~\cite{LS5}, in which isospectral reduction onto three
distinguished sites reveals a higher-dimensional latent symmetry that
is not manifest in the geometry of the full network. If these sites
were continuously monitored, the corresponding measurement-selected
Zeno sectors could form a symmetry-related family, while the
complementary unmonitored sector would retain a higher-dimensional
internal dynamics. This example illustrates how the mechanism explored
here could extend naturally to more complex networks, multiple
monitored sites, and higher-dimensional latent-symmetry structures.

Our results thus extend the role of latent symmetry from spectral,
transport, and topological phenomena to monitored quantum dynamics.
This connection may also be relevant to quantum technologies in which
measurement and dynamical constraints are active resources, including
quantum-state control, quantum simulation, and the engineering of
protected dynamical subspaces. More generally, latent symmetry could
provide a design principle for monitored quantum networks, enabling
symmetry-constrained dynamics without imposing the corresponding
geometrical symmetry on the physical device.

Several directions naturally follow from the present work. Extending
the analysis to multiple excitations and interactions could reveal
latent-symmetry-constrained many-body Zeno dynamics, while collective
monitoring, engineered dissipation, and imperfect detection would
broaden the class of measurement-induced subspaces that can inherit
hidden symmetry constraints. Conversely, trajectory statistics could
be explored as an operational probe of latent symmetries that are not
known a priori from the microscopic structure of the system.\\
\\

\noindent
\textbf{Acknowledgments}\\
This work was supported by grant CEX2021-001164-M (Maria de Maeztu Program for
Units of Excellence in R\&D) funded by MICIU/AEI/10.13039/501100011033..\\
\\
\noindent
\textbf{Conflict of Interest}\\
The author declares no conflict of interest.\\
\\
\noindent
\textbf{Data Availability Statement}\\
The data that support the findings of this study are available from the corresponding author upon reasonable request.\\
\\
\noindent
\textbf{Keywords}\\
latent symmetry, quantum Zeno dynamics, continuous quantum
measurement, quantum trajectories, circuit quantum electrodynamics.

\end{document}